# From Spin States to Social Consensus: Ising Approach to Dimer Configurations in Opinion Formation

Yasuko Kawahata [†]

Faculty of Sociology, Department of Media Sociology, Rikkyo University, 3-34-1 Nishi-Ikebukuro,Toshima-ku, Tokyo, 171-8501, JAPAN.
`ykawahata@rikkyo.ac.jp,kawahata.lab3@damp.tottori-u.ac.jp`

**Abstract:** The field of opinion dynamics has evolved steadily since the earliest studies applying magnetic physics methods to better understand social opinion formation. However, in the real world, complete agreement of opinions is rare, and biaxial consensus, especially on social issues, is rare. To address this challenge, Ishii and Kawabata (2018) proposed an extended version of the Bounded Confidence Model that introduces new parameters indicating dissent and distrust, as well as the influence of mass media. Their model aimed to capture more realistic social opinion dynamics by introducing coefficients representing the degree of trust and distrust, rather than assuming convergence of opinions. In this paper, we propose a new approach to opinion dynamics based on this Trust-Distrust Model (TDM), applying the dimer allocation and Ising model. Our goal is to explore how the interaction between trust and distrust affects social opinion formation. In particular, we analyze through mathematical models how various external stimuli, such as mass media, third-party opinions, and economic and political factors, affect people's opinions. Our approach is to mathematically represent the dynamics of trust and distrust, which traditional models have not addressed. This theoretical framework provides new insights into the polarization of opinions, the process of consensus building, and how these are reflected in social behavior. In addition to developing the theoretical framework by applying the dimer configuration, the dimer model and the Ising model, this paper uses numerical simulations to show how the proposed model applies to actual social opinion formation. This research aims to contribute to a deeper understanding of social opinion formation by providing new perspectives in the fields of social science, physics, and computational modeling.

**Keywords:** Toroidal Structure,Dimer Configurations, Ising Models, Opinion Dynamics, Social Polarization, Consensus Formation, Network Analysis, Mathematical Modeling, Spin Systems, Complex Systems, Computational Social Science

## 1. Introduction

In recent years, the study of opinion dynamics has gained significant attention, particularly in the context of understanding social phenomena, media influence, and consensus formation. Traditional models, such as the Ising model and dimer configurations, have provided valuable insights into the mechanisms underlying these processes. This paper seeks to explore new approaches to opinion dynamics, leveraging these classic models to address contemporary challenges. Opinion dynamics theory has undergone various transformations over the years. Early studies, such as those by Serge Galam (1982), employed techniques from magnetic physics to simulate binary opposition in opinion formation, namely support and opposition, or support and ignorance.

These models were instrumental in unraveling the complexities of social interactions and the formation of collective opinions. Since 2000, there has been a shift towards analyzing opinions as continuously varying quantities, rather than binary states. This transition is exemplified by the Bounded focus on how trust and distrust shape the landscape of social opinion and its consequences on societal structures. We argue that understanding the interplay of trust and distrust is essential for grasping the full spectrum of opinion dynamics in contemporary society. Modeling human behavior in social contexts requires an understanding of the stimuli influencing individual actions. This paper posits that human behavior is influenced by a range of external stimuli, including media information, economic factors, political motivations, and personal relationships. We propose a mathematical model that captures these influences, providing a framework for analyzing the complex nature of human interactions in society. This paper presents a novel approach to opinion dynamics, integrating the principles of the Ising model and dimer configurations with the contemporary Trust-Distrust Model. Our objective is to offer new perspectives on understanding the multifaceted nature of opinion formation and its implications for society.

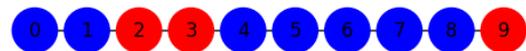

Fig. 1: Network Graph Based Ising Model

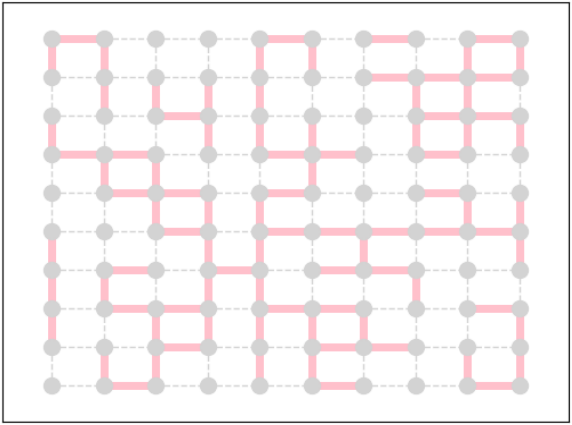

Fig. 2: Network Graph Based Dimer Tiling Map

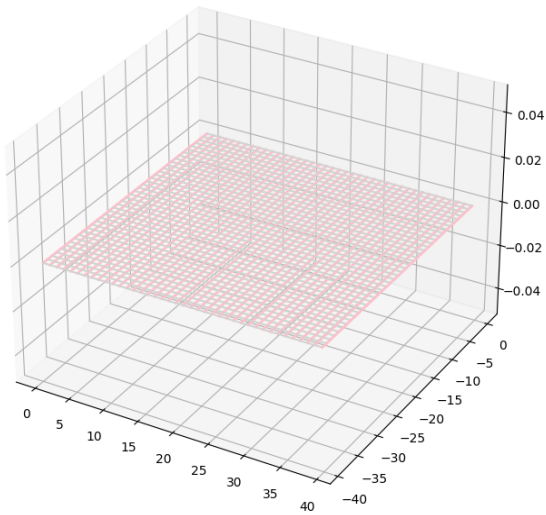

Fig. 3: Dimer Tiling Map to Torus Grid Representation

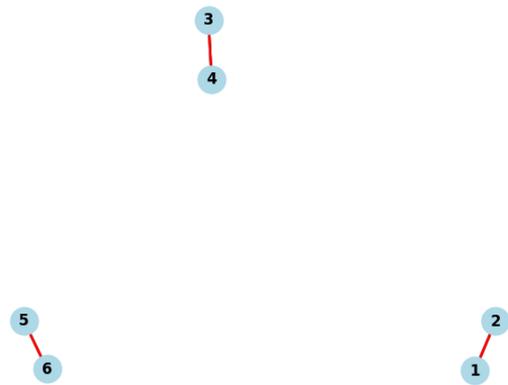

Fig. 4: Based on Dimer Positions

Confidence Model, which presents a more nuanced approach to understanding opinion dynamics. Pioneered by researchers like Gérard Weisbuch, Guillaume Deffuant, and Hegselmann-Krause (2002), this model introduced the concept of influence coefficients, confined to positive values, which significantly impact the speed of opinion convergence. The Trust-Distrust Model (TDM), developed by Ishii and Kawahata (2018), extends the Bounded Confidence Model by incorporating parameters that represent opposition, distrust, and external influences such as mass media. Crucially, this model does not limit coefficients to positive values, allowing for negative values that signify distrustful relationships. This approach provides a more realistic representation of social dynamics, where consensus is not always the outcome, and divergent opinions can persist. This paper aims to explore various social simulation case studies using the TDM. In particular, we

### 1.1 Unveiling the Complexities of Opinion Dynamics: A Dimer-Ising Model Analysis

The process of social opinion formation is inherently complex and constituted by a variety of factors. One powerful approach to understanding this complexity is the application of mathematical models borrowed from physics. In this paper, we propose a novel approach combining dimer configurations with the Ising model, applying it to the study of opinion dynamics.

The Ising model, widely used in physics, was developed to simulate spin interactions in magnetic systems. It is based on the simple assumption that each spin takes one of two states: +1 (up) or -1 (down). The mathematical expression of the Ising model is as follows:

$$H = -J \sum_{\langle i,j \rangle} s_i s_j$$

where $H$ is the system's Hamiltonian, $J$ represents the interaction strength between adjacent spins, $s_i$ denotes the spin state, and the sum is over all pairs of adjacent spins.

On the other hand, dimer configurations, used in combinatorics and statistical physics, can be applied in the context of opinion dynamics to analyze patterns of opinion distribution and change. Dimers represent situations where two adjacent elements pair up, and this can be used to represent interactions and equilibrium states of opinions.

In this study, we apply the Ising model to opinion dynamics, using dimer configurations to capture the dynamics of opinion distribution and interaction. Specifically, we model the process of opinion polarization and consensus formation by linking changes in spin states with updates in dimer configurations. This approach aims to deepen our understanding of opinion dynamics and elucidate the mechanisms of social opinion formation.

Ultimately, this paper aims to provide a new theoretical framework and mathematical approach to opinion dynamics, potentially offering a fresh perspective in research on opinion formation in social sciences, political science, and economics.

## 2. Dimer Configuration and Dimer Model

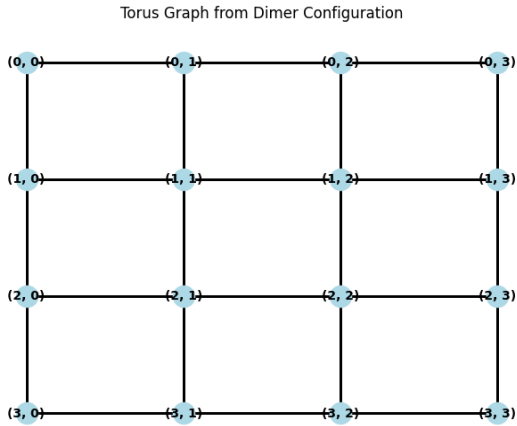

Fig. 5: Torus graph Based on Dimer Positions

We will provide explanations about dimer configurations and the dimer model, including mathematical expressions. The dimer model is mainly used in statistical physics and condensed matter physics to study various arrangements where adjacent sites on a lattice are paired (bonded as dimers).

### 2.1 Dimer Configuration

In a dimer configuration, each edge on a lattice (or graph) is occupied by at most one dimer. In this case, each vertex belongs to only one dimer. For example, in a 2D square lattice, a dimer configuration is represented by whether each edge of the lattice is occupied by a dimer or not.

Mathematically, a dimer configuration $\mathcal{D}$ is represented as a function $D : E \to \{0, 1\}$ defined on the edge set $E$. Here, $D(e) = 1$ means that edge $e$ is occupied by a dimer, while $D(e) = 0$ means it is unoccupied.

### 2.2 Dimer Model

In the dimer model, all possible dimer configurations on the lattice are considered. Generally, the partition function $Z$ of the dimer model is defined as the sum over all possible dimer configurations $\mathcal{D}$, where the energy associated with a specific dimer configuration is denoted as $E(\mathcal{D})$:

$$Z = \sum_{\mathcal{D}} e^{-\beta E(\mathcal{D})}$$

Here, $\beta = \frac{1}{kT}$ is the inverse temperature parameter, where $k$ is the Boltzmann constant and $T$ is the absolute temperature.

### 2.3 Statistical Mechanics of Dimers
### 2.4 Applications of Dimers

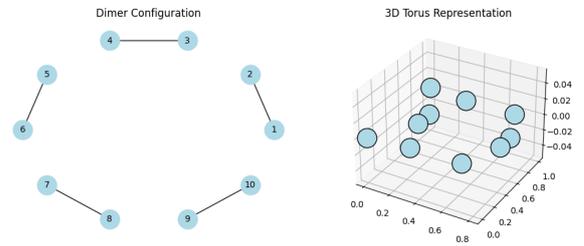

Fig. 6: Torus Representation, Based on Dimer Positions

### 2.5 Applications of Dimers

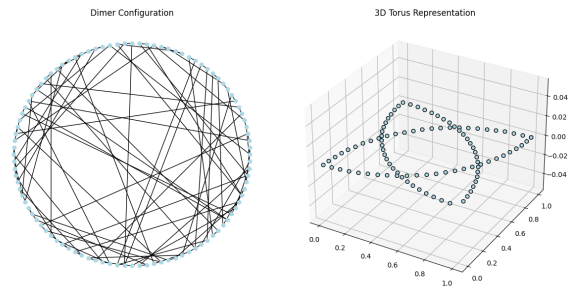

Fig. 7: Torus Representation, Based on Dimer Positions

The statistical mechanical properties of the dimer model are analyzed using the partition function $Z$. For example, the system's free energy $F$ is given by $F = -kT \ln Z$. Additionally, the probability distribution of dimer configurations is given as a normalized weight using the partition function:

$$P(\mathcal{D}) = \frac{e^{-\beta E(\mathcal{D})}}{Z}$$

The dimer model finds applications in a wide range of physical and mathematical problems, including phase transitions, topological phases, and quantum algorithm design for quantum computers. Particularly, the behavior of dimers at low temperatures and their topological properties are essential topics in condensed matter physics.

## 2.6 Points

- The dimer model is generally challenging to handle analytically and often relies on numerical simulations. - The mathematical treatment of dimer configurations can vary significantly depending on the type of lattice and its dimension.

# 3. Dimer Coverings: Definition and Basics

we provide a detailed explanation of Dimer coverings from the perspective of group theory. We begin by understanding how group-theoretic operations are applied to Dimer coverings. While the characteristics of Dimer coverings themselves do not have a direct relationship with group theory, group theory plays a crucial role in analyzing their symmetry.

Dimer coverings are defined by a function on the edge set $E$, represented as $D : E \to \{0, 1\}$. Here, $D(e) = 1$ indicates that edge $e$ is occupied by a dimer, while $D(e) = 0$ indicates that it is unoccupied. This definition ensures that each vertex belongs to at most one dimer.

## 3.1 Application of Group-Theoretic Operations

To analyze the symmetry of Dimer coverings, group-theoretic operations are employed. These operations consist of rotation ($R$), reflection ($M$), and translation ($T$), which help determine whether a Dimer covering remains invariant under specific symmetry operations.

## 3.2 Analysis of Symmetry

The analysis of symmetry using group theory involves the following steps:

(1) **Definition of the Group ($G$):** Define a group $G$ that includes operations such as rotation, reflection, translation, etc.

(2) **Application of Group Operations:** Apply each element $g$ of $G$ to the Dimer covering $D$ to obtain a new configuration $g(D)$.

(3) **Verification of Invariance:** Check whether $g(D)$ is equal to $D$; if it is, then $D$ is considered invariant under the operation $g$.

## 3.3 Point of view

As an example, consider a 2D square lattice. If the group $G$ includes 90-degree rotations, you would verify whether each Dimer covering $D$ remains invariant under this rotation. The steps are as follows:

(1) Define the rotation operation $R_{90}$.

(2) Apply $R_{90}$ to $D$ to obtain a new configuration $R_{90}(D)$.

(3) Check if $R_{90}(D)$ is equal to $D$.

By repeating this process for all elements of group $G$, you can determine the invariance of $D$ under different symmetry operations. When considering Dimer configurations in quotient spaces and their symmetry, the following concepts are essential:

Transform a 2D square lattice into a quotient space using the action of a group $G$, such as $C_4$ (the 90-degree rotation group). Quotient spaces classify points from the original space into equivalence classes. A point $[p]$ in the quotient space represents the point $p$ in the original space and the set of all points moved to $p$ by the action of $G$.

## 3.4 Representation of Dimer Configurations

Dimer configurations on the quotient space $\mathcal{D}$ are represented as a function $\tilde{D} : \tilde{E} \to \{0, 1\}$ on the equivalence class set of edges $\tilde{E}$. Specifically:

$$\tilde{D}([\tilde{e}]) = \begin{cases} 1 & \text{if the edge equivalence class } [\tilde{e}] \text{ is occupied by a dimer} \\ 0 & \text{otherwise} \end{cases}$$

In this section's conclusion, group theory provides a powerful framework for analyzing the symmetry of Dimer coverings. By understanding how group-theoretic operations act on Dimer configurations and considering quotient spaces, we can gain valuable insights into the invariance properties of Dimer coverings under various symmetries.

The one-to-one correspondence between periodic domino tilings and the ground state configurations of the 2D periodic lattice Ising model with full frustration implies that both possess equivalent physical and mathematical structures. To explain this mathematically, we first need to understand the fundamental characteristics of each model.

## 3.5 Tilings: Basic Concepts with Ising Model

Tilings refer to arrangements that completely cover a 2D lattice using pieces, where each domino can cover two adjacent lattice points. One crucial characteristic of tilings is that no two dominoes overlap. The Ising model consists of lattice points where spins can take either an up (+1) or down (-1) orientation. In the case of the fully frustrated Ising model, there is a negative interaction between all neighboring spin pairs, making it desirable for adjacent spins to have opposite orientations.

## 3.6 Applications of Dimers

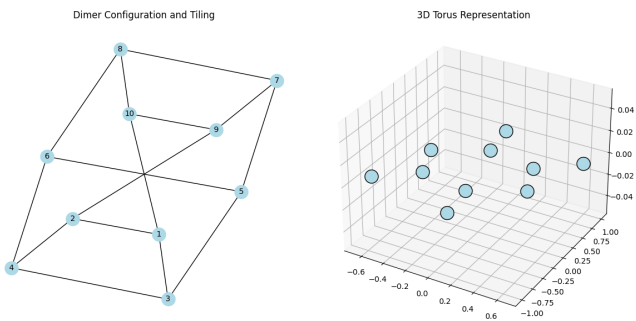

Fig. 8: Torus Representation, Based on Dimer Positions Tilimg

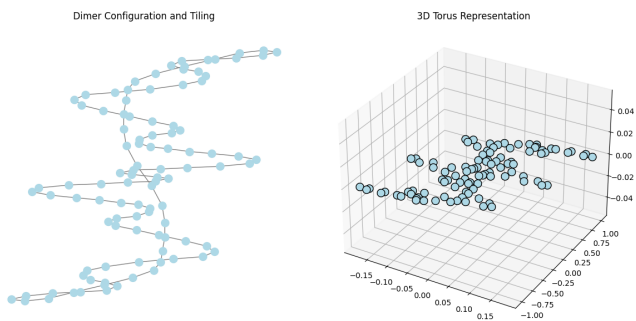

Fig. 9: Torus Representation, Based on Dimer Positions Tilimg

## 3.7 One-to-One Correspondence

The correspondence between domino tilings and the frustrated Ising model implies that domino configurations can be directly mapped to the spin configurations of the Ising model. Specifically:

1. Domino Orientation and Spin Orientation: The orientation of each domino in domino tilings (horizontal or vertical) corresponds to the orientation of neighboring spins (up or down) in the Ising model.

2. Minimization of Interaction:** When dominoes cover adjacent lattice points, in the Ising model, this corresponds to adjacent spins adopting opposite orientations. In the fully frustrated Ising model, there is a need to minimize the energy of all neighboring spin pairs.

## 3.8 Mathematical Representation

The energy function of the Ising model is represented as follows:

$$H = -J \sum_{\langle i,j \rangle} s_i s_j$$

Here, $s_i$ and $s_j$ represent neighboring spins, $J$ is the interaction constant (negative in the frustrated model), and $\langle i, j \rangle$ denotes neighboring spin pairs.

In domino tilings, the rule that each domino covers two adjacent lattice points corresponds to the rule in the Ising model that adjacent spins must have opposite orientations.

## 3.9 Considerations

Interpreting the relationship between periodic domino tilings and the fully frustrated 2D lattice Ising model in the context of opinion dynamics is feasible. This approach involves using opinion dynamics models to interpret the spin configurations of the Ising model as "opinions" and studying their interactions and dynamics.

## 3.10 Relationship with Opinion Dynamics

In the Ising model, each spin can be viewed as representing either "approval" or "disapproval" of an opinion, with +1 and -1 corresponding to these states, respectively. In the fully frustrated Ising model, the desired state is for neighboring spins (opinions) to be in opposition, which can be interpreted as opinion polarization.

## 3.11 Domino Tilings and Opinion Patterns

Domino tilings can be seen as patterns of opinions. By placing dominoes, patterns are created where adjacent opinions (spins) are in opposition. This pattern can represent equilibrium states or polarization states of opinions in a social network. Equilibrium and Polarization of Opinions:The ground states of the fully frustrated Ising model can be interpreted through opinion dynamics as states of opinion equilibrium or extreme polarization. In social networks, the presence of opposing opinions among neighbors can lead to tension or conflict. Opinion Dynamics:Interpreting the dynamics of the Ising model as opinion changes allows us to explore how opinions evolve over time and reach new equilibrium states. Model Limitations:This interpretation assumes similarities between the Ising model and opinion dynamics, but real-world social opinion formation is influenced by more complex and diverse factors. Understanding the limitations of the model is essential. Interpreting the relationship between the Ising model, domino tilings, and opinion dynamics may provide valuable insights into social interactions and the opinion formation process. However, modeling real social phenomena requires consideration of more intricate elements.

## 3.12 Mathematical Model of Opinion Dynamics Using Dimer Configurations

Dimer configurations are generated based on the spin states $s_i$ of the Ising model, where $s_i \in \{-1, +1\}$.

The update rule for spin states is given as follows:

$$s_i^{(t+1)} = \text{sign}\left(\sum_{j \in \text{neighbors}(i)} s_j^{(t)}\right)$$

The update condition for dimer configurations is as follows:

$$D_{ij}^{(t+1)} = \begin{cases} 1 & \text{if } s_i^{(t+1)} \neq s_j^{(t+1)} \\ 0 & \text{otherwise} \end{cases}$$

### Merits

**Simplicity**: The model is represented by the following simple equation, making computations easy:

$$s_i^{(t+1)} = \text{sign}\left(\sum_{j \in \text{neighbors}(i)} s_j^{(t)}\right)$$

**Intuitive Understanding**: Spin state flips mimic changes in societal opinions and are intuitively easy to understand.

**Local Interactions**: Each spin is influenced only by its local neighboring spins, eliminating the need to consider complex network structures.

### Demerits

**Excessive Simplification**: Real-world societal opinion formation is more complex and may not be fully captured by the following equation:

$$D_{ij}^{(t+1)} = \begin{cases} 1 & \text{if } s_i^{(t+1)} \neq s_j^{(t+1)} \\ 0 & \text{otherwise} \end{cases}$$

**Limitations in Dynamics**: The model considers changes over time but has limitations in capturing long-term dynamics or external influences.

**Lack of Diversity**: Spin states are binary, whereas real opinions are more diverse, making it unable to represent certain scenarios.

## 3.13 Considerations Regarding Social Phenomena

### Understanding Polarization

- The model based on dimer configurations allows for the modeling of polarization in societal opinions. By likening the spin states in the Ising model to societal opinions, we can observe how opinions divide among groups in conflict.
- Equilibrium of Opinions: Dimer configurations on a torus can represent equilibrium states of opinions in a social network. It enables the study of how opinions change over time and reach new equilibrium states. - Challenges: The dimer configuration model cannot fully capture the complexity and diversity of real social phenomena. Real-world social phenomena are influenced by many different factors and are often more dynamic and unpredictable.

## 3.14 Consideration as Media Influence

- Role of Media: The dimer configuration model allows us to consider the influence of media on societal opinion formation. It can model how information provided by the media affects the formation and change of opinions. - Diffusion of Information: The diffusion of information through the media is similar to the propagation of opinions in a network. This allows us to analyze how information spreads rapidly and influences societal opinions. - Challenges: The real media environment is highly diverse, and explaining its influence with a single model is difficult. Additionally, issues such as media bias and misinformation need to be considered.

## 3.15 Expectations and Challenges in Consensus Formation

- Promoting Consensus: Using the dimer configuration model, we can observe the consensus formation process among individuals with different opinions. It allows us to analyze how conflicts of opinions are resolved and consensus is reached. - Diversity of Opinions: The model can be used to study opinion exchange and its impact among individuals with diverse opinions. It explores the possibilities of integrating and coexisting different perspectives. - Challenges: Consensus formation is a highly complex process, and the model may not capture all elements. Especially in real social environments, individual emotions and cultural backgrounds can have a significant impact.

## 4. Considerations Based on the Features of the Opinion Dynamics Model Using Dimer Configurations

### 4.1 Features of the Model Equations

The model equations are based on the interaction of local spin states (opinions). This interaction is expressed as follows:

$$s_i^{(t+1)} = \text{sign}\left(\sum_{j \in \text{neighbors}(i)} s_j^{(t)}\right)$$

$$D_{ij}^{(t+1)} = \begin{cases} 1 & \text{if } s_i^{(t+1)} \neq s_j^{(t+1)} \\ 0 & \text{otherwise} \end{cases}$$

Here, $s_i$ represents the spin (opinion) of node $i$, and $D_{ij}$ represents dimer configurations (opinion equilibrium).

## 4.2 Considerations Regarding Social Phenomena

**Influence of Local Interactions**: The model equations suggest that individual opinions are influenced only by direct social contact. This local interaction is key to understanding the propagation of opinions and polarization within a social network.

**Stabilization of Opinions**: The model implies a process where opinions stabilize over time. This stabilization may lead to social equilibria and polarization states.

## 4.3 Consideration as Media Influence

**Modeling Information Diffusion**: Media-induced information diffusion can be considered as local interactions modeled by the equations. This makes it easier to understand the impact of media on opinion formation.

**Limits and Impact of Media**: The influence of media is limited to local interactions in this model. In actual society, media has broader influence and plays a significant role in opinion equilibria and polarization.

## 4.4 Expectations and Challenges in Consensus Formation

**Towards Consensus Formation**: The model equations can illustrate the consensus formation process among individuals with different opinions. It is particularly useful for studying the process of resolving conflicts of opinions.

**Opinion Dynamics and Diversity**: While the model captures opinion dynamics, real consensus formation involves cultural, emotional, or personal factors. This diversity is not fully represented by the model equations.

## 5. Discussion

### 5.1 Analysis of Network Graph Based on Dimer Positions: An Experimental Approach

**Formulation and Parameters of the Model**

The following formulation describes the process of constructing a network graph from dimer positions, utilizing concepts from K-means clustering and the Erdős–Rényi model, followed by an exploration of the network through Breadth-First Search (BFS).

**Node Opinion Generation**: Each node $i$ is assigned an opinion $O_i$, generated randomly within the range from 0 to 1.

$$O_i \sim \text{Uniform}(0, 1), \quad i = 1, 2, \ldots, \text{num\_nodes}$$

**K-means Clustering**: The opinions are classified into $k$ clusters, where $k = 3$.

$$\text{KMeans}(\text{opinions}, k = 3)$$

Each node $i$ is assigned a label $\text{labels}_i$ based on the clustering result.

**Network Generation**: A network is generated using the Erdős–Rényi model, with number of nodes as num\_nodes = 10 and edge existence probability $p = 0.1$.

$$G = \text{ErdosRenyiGraph}(\text{num\_nodes}, p = 0.1)$$

**Network Visualization**: The network is visualized with nodes colored based on the clustering results.

**BFS Execution and Visualization**: BFS is executed from a specified node start\_node, and the results are visualized.

$$\text{BFSTree}(G, \text{start\_node})$$

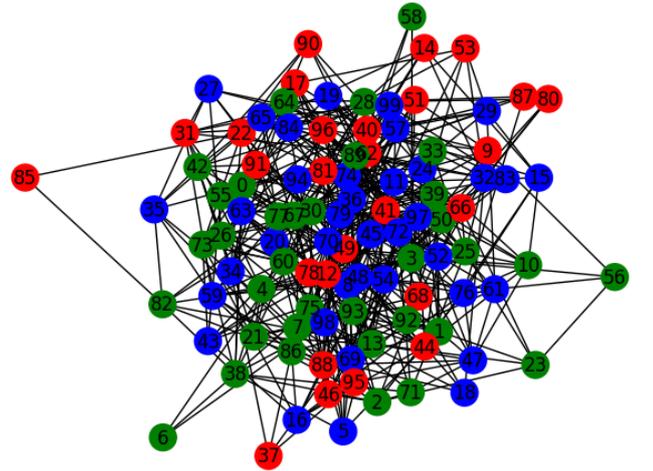

Fig. 10: Network Graph Based on Dimer Positions

Based on the Erdős–Rényi model and the utilization of K-means clustering and Breadth-First Search (BFS) in the provided network graphs, we can extend the analysis to include the consideration of dimer positions and their implications, in addition to the social phenomena, media influence, and consensus formation previously discussed. Here's an analysis considering all four points:

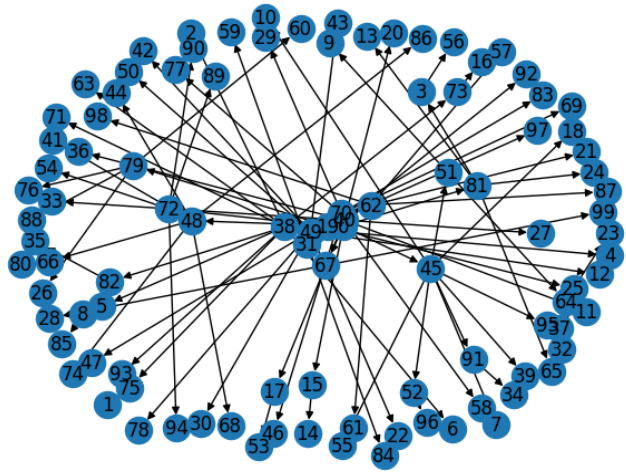

Fig. 11: Network Graph Based on Dimer Positions

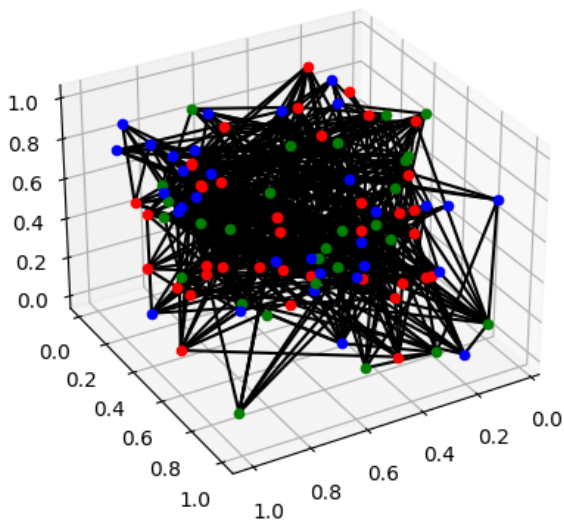

Fig. 12: Network Graph Based on Dimer Positions

### 1. Social Phenomenon

The clustering in the network could be indicative of social groups with shared opinions or interests. The existence of tightly knit clusters may suggest strong in-group relationships, potentially leading to echo chambers where similar views are reinforced. Sparse connections between these clusters could indicate limited interaction between different social groups, possibly leading to polarization.

### 2. Media Influence

The impact of media on the network could be inferred from the centrality of certain nodes. If nodes representing media sources are centrally located and have many connections, this suggests a significant influence on the network's opinion dynamics. The media's role in shaping opinions might be seen in the formation of new clusters or the strengthening of existing ones.

### 3. Consensus Formation

Consensus within the network could be seen when a majority of nodes within a cluster share the same opinion, and there is a significant spread of this opinion to other parts of the network through BFS pathways. The speed and reach of this spread can indicate how quickly a consensus can be formed across the network.

### 4. Dimer Position Trends

The dimer positions, which might represent paired connections or relationships between nodes, could provide insight into the underlying structure and dynamics of the network. High-frequency dimer positions could indicate common pathways for information flow or shared opinion dynamics. The distribution of these dimers, whether centralized or dispersed, might suggest either a few influential nodes or a more distributed influence across the network.

When analyzing dimer positions, we could consider factors such as the frequency of dimer occurrences, the diversity of connections, and their distribution across the network. This can reveal the network's resilience to change or its vulnerability to the spread of misinformation. A network with diverse and widespread dimer positions may suggest robustness, while one with centralized dimers might be more susceptible to targeted interventions.

Overall, these network graphs and their dimer positions can offer a multi-faceted view of a system's connectivity and dynamics, allowing for a deeper understanding of how entities within the network interact and influence one another. To conduct a more detailed analysis, it would be necessary to understand the specific real-world entities these nodes and dimers represent, as well as the nature of the data and the context of the network's application.

## 5.2 Generation of Tiling (`generate_tiling` function)

Grid Size: $N$ (in the code, `grid_size`)

Probability of Tile Placement: $p$ (in the code, `tile_prob`)

The generated graph $G$ is an $N \times N$ grid graph.

For each edge $(u, v)$ in $G$, a tile is placed with probability $p$:

if np.random.rand() < $p$, then $G[u][v]['tiled'] = $ True

Otherwise, no tile is placed:

else $G[u][v]['tiled'] = $ False

### Visualization of Tiling (`plot_tiling` function)

Edges with tiles are drawn in blue ('blue'), while others are in red ('red').

Node positions (`pos`) in the graph are mapped from $(x, y)$ to $(y, -x)$.

### Analysis of Tiling (`analyze_tiling` function)

Count the number of tiled edges $T$ in the graph:

$T = \text{len}(\{e \text{ for } e \text{ in } G.\text{edges}() \text{ if } G[e[0]][e[1]]['tiled']\})$

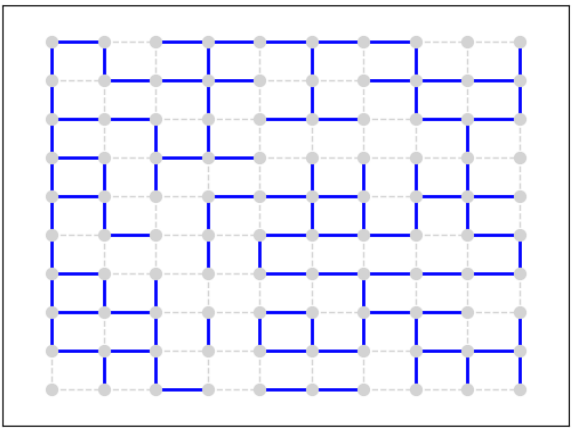

Fig. 13: Network Graph Based on Dimer Positions

### 5.3 Analysis of Dimer Tiling in Various Contexts

This approach is particularly relevant in the study of perfect matchings or dimer coverings in graph theory.

### 1. Social Phenomenon

In a social context, the dimer configurations could represent binary relationships or partnerships within a network. A perfect dimer covering, where every node is part of exactly one dimer, might illustrate a society with an ideal balance of relationships or resources. If the tiling represents social bonds, the complete coverage might suggest a highly interconnected community without isolated individuals.

### 2. Media Influence

If we treat dimers as channels of communication, the tiling can be a metaphor for the spread of information through a network where each dimer represents a medium through which information flows. In this model, the uniformity of the tiling could signify equal access to information across the network, with no single node being more influenced or informed than another due to the regularity of connections.

### 3. Consensus Formation

Consensus formation might be visualized in this context as the process by which information or agreement spreads through these dimer connections. If we assume that a consensus is reached when information has flowed through all available paths, then this tiling, with its complete and regular dimer placement, could represent a network where consensus is quickly achieved due to the uniformity and efficiency of connections.

### 4. Trends in Dimer Positioning

The image shows a regular pattern, suggesting a highly structured approach to dimer placement. This uniformity could indicate robustness in the network's structure, where the removal of a dimer doesn't disrupt the system due to the presence of many alternative pathways. However, it might also indicate a lack of flexibility or adaptability in the face of changing conditions, as the system is highly ordered and may rely on the maintenance of that order.

In the realms of physics and mathematics, such a tiling might be used to explore properties like entropy, phase transitions, or to solve problems related to the enumeration of perfect matchings. In more applied fields like sociology or economics, analogous structures could be used to model optimal pairings or allocations of resources. To provide a deeper analysis, one would need more context about the specific application or the system that this dimer tiling is intended to model.

## 6. Conclusion

Finally, the dimers are randomly placed on the grid, numerically calculated using the Ising model, and then the network diagram is re-expanded to visualize them on the torus. A coordinate transformation of the torus is used to map the position of each node on the grid to the 3D coordinates of the torus. We expect that such numerical experiments will be useful for visualizing topological data that take into account spatio-temporal information on the opinion dynamics and for understanding complex network structures.

## Generation of Dimer Configurations (`create_dimer_configuration` function)

Grid Size: $N$ (in the code, `grid_size`)

The generated graph $G$ is an $N \times N$ grid graph with periodic boundary conditions.

Dimer placement on each edge $(u, v)$ is randomly assigned:

$$G[u][v]['dimer'] = \text{np.random.choice}([0, 1])$$

where 0 represents no dimer and 1 represents a dimer.

## Torus (`plot_torus_network` function)

Torus coordinates for each node are computed as follows:

$$x = \frac{\text{pos}[u][0]}{N} \times 2\pi, \quad y = \frac{\text{pos}[u][1]}{N} \times 2\pi$$

$$z = \cos(x), \quad r = 2 + \sin(y)$$

$$x = r \times \cos(x), \quad y = r \times \sin(y)$$

where $x, y$ are the coordinates on the grid, and $r$ is the radius of the torus.

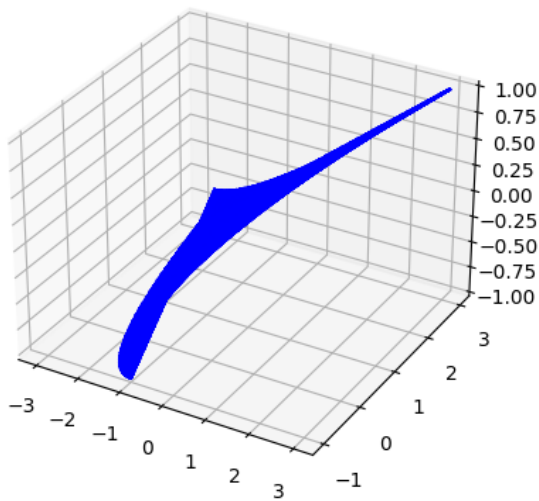

Fig. 14: Ising Network Graph Based on Dimer Positions

Results provided appears to represent a result from a numerical experiment involving the Ising model for dimer placement on a torus and the visualization of the network. The Ising model is a mathematical model of ferromagnetism in statistical mechanics, and its adaptation to dimer placement can be complex. In this model, the system's energy states can be mapped to social, informational, or consensus states within a network.

### 1. Social Phenomenon

The structure shown might represent a social network with complex interactions. For example, the path could represent the flow of social influence through the network, where the width or trajectory could indicate the strength and direction of this influence. In a society, this could visualize segregated communities or the spread of a social trend.

### 2. Media Influence

If the plot represents different media influences over time and space, the broadening or narrowing of the path could signify the varying impact of media on public opinion. For instance, a wider path may indicate a strong, unified media influence, while a narrow one may suggest a fragmented media landscape with diverse and possibly conflicting information sources.

### 3. Consensus Formation

The plot could be interpreted as the process of reaching a consensus in a population. A consensus might be reached quickly in areas where the path is direct and unobstructed, but more slowly or not at all in areas where the path is twisted or disjointed.

### 4. Reconstruction from Dimer Network to Torus

The shape may represent the topological transformation of a dimer-covered network into a toroidal structure. In this context, it could reflect the complexities of mapping local interactions (dimers) onto a global structure (torus), which could be an analogy for understanding how local rules or behaviors can give rise to global patterns or structures in a society.

### 5. Dimer Positioning

In the context of the dimer model, the positioning of dimers could represent the fundamental connections or relationships within the network. The way these connections are visualized in 3D space on a torus could provide insights into the robustness of the network, the presence of critical links, or the potential for network disruption. These interpretations are speculative and would require a deeper understanding of the underlying data and the parameters of the numerical experiment to be precise. In practical terms, such a visualization can help researchers understand the behavior of complex systems, whether they be social networks, magnetic materials, or other systems where the Ising model and network theory are applicable.

# Aknowlegement

The author is grateful for discussion with Prof. Serge Galam and Prof.Akira Ishii. This research is supported by Grant-in-Aid for Scientific Research Project FY 2019-2021, Research Project/Area No. 19K04881, "Construction of a new theory of opinion dynamics that can describe the real picture of society by introducing trust and distrust".